\documentclass[draftcls,twocolumn,twoside]{IEEEtran}
\usepackage[]{authblk}
\usepackage{amsmath,amstext,amsthm,amssymb,epsf}
\usepackage{latexsym} 
\usepackage[pdftex]{graphicx}
\graphicspath{ {./figures} {./figures/ac_figure_1}}
\usepackage{setspace} 
\usepackage{float}
\usepackage[section]{placeins}
\usepackage[]{hyperref}
\usepackage{caption} % hypcap is true by default so [hypcap=true] is optional in \usepackage[hypcap=true]{caption}
\usepackage{xcolor}
\usepackage[]{soul}
\usepackage{siunitx}  % For the 'S' column type
\usepackage{booktabs} % For \toprule, \midrule, \bottomrule
\usepackage[bibstyle=nature,citestyle=nature]{biblatex}
\usepackage{graphicx}
\usepackage{multirow}
\usepackage{setspace} % Controls line spacing safely

\hypersetup{
    colorlinks=true,
    linkcolor={blue},
    citecolor={blue!50!black}, 
    urlcolor={blue!80!black}
}
\numberwithin{equation}{section} % in amsmath

\newtheorem{coro}{\sc Corollary}
\newtheorem{req}{\sc Requirement}

\newtheorem{defin}{\sc Definition}
\newtheorem{rem}{\sc Remark}
\newtheorem{cla}{\sc Claim}
\newtheorem{ex}{\sc Example}

\begin{document}
\onehalfspacing 
\title{Algorithmic statistics of retinal images}

\author[1]{Loan Huynh}
\author[2,3]{Ronald Zambrano}
\author[1,2]{Layton Aho}
\author[2,5]{Fabio Lavinsky}
\author[2,4,5]{Gadi Wollstein}
\author[2,3,4,5]{Joel S. Schuman}
\author[1]{Andrew R. Cohen}

\affil[1]{Department of Electrical and Computer Engineering, Drexel University, Philadelphia, PA, USA}
\affil[2]{Glaucoma Service, Wills Eye Hospital, Philadelphia, PA, USA}
\affil[3]{School of Biomedical Engineering, Science and Health Systems, Drexel University, Philadelphia, PA, USA}
\affil[4]{Vickie and Jack Farber Vision Research Center, Wills Eye Hospital, Philadelphia, PA, USA}
\affil[5]{Sidney Kimmel Medical College, Thomas Jefferson University, Philadelphia, PA, USA}
\affil[*]{correspondence to andrew.r.cohen@drexel.edu}
\maketitle

% pandoc RSF.tex --citeproc --bibliography=rsf.bib --csl=nature.csl -o RSF.docx

\section{Abstract}
There has been a tremendous amount of image processing and machine learning research to measure and classify disease progression from live optical coherence tomography (OCT) imaging of the retina. The images considered here are large, complex, three-dimensional (3-D) and difficult to visualize effectively. Many current supervised machine learning approaches, \emph{e.g.} neural networks, are non-metric meaning that any features or measurements generated can introduce systematic distortion that may be correlated with underlying non-meaningful physiological differences. Here we present a metric learning approach using the normalized compression distance (NCD) combined with anisotropic structure-enhancing filters to quantify and visualize the principal differences among a collection of 3-D retinal images. We validate the NCD-measured structural differences between pairs of images against the physician-measured change in visual field function, achieving a prediction error of $\sim$ 0.5 dB, more accurate than non-metric deep learning approaches. The normalized compression vectors (NCV) are proposed as a feature set measuring visual differences among a collection of 3-D microscopy images. The utility of the NCV for visualizing and measuring patterns of change is demonstrated for a human with moderate non-progressing glaucoma and for a non-human primate model using intraocular pressure setting manipulation. We conclude with a brief simulation of non-metric embedding features, \emph{e.g.} from neural networks, introducing class-correlated statistical distortion.

\section{Introduction}
In vivo images of retinal tissue are a critical source of information for understanding the structure and function of the retina, for diagnosing and monitoring diseases such as glaucoma and also for other central nervous system-related diseases \cite{schizophrenia}. The relationship between retinal structure measured from images and function measured in the ophthalmology clinic is complex and non-linear, making it difficult to predict functional changes from structural changes. \textbf{In this work, we propose a novel approach to quantifying and visualizing changes in 3-D \emph{in vivo} retinal images using the normalized compression distance (NCD) \cite{Vitanyi2005} combined with multi-resolution anisotropic scale-space filters \cite{HIP}}. The NCD is parameter-free and uses lossless 3-D image compression to compute a metric distance between pairs of images \cite{Vitanyi2008,Vitanyi2005,Cohen2015_NCDM,aho}. We combine the NCD with the multi-scale structure-enhancing filters as in \cite{frangi,lindeberg1994scale} to quantify the visual differences between images at different spatial scales. The scale-space filters allow us to concisely query structures at different spatial resolutions, forming the parameterized visual input to the NCD. Common scale-space implementations include the Gaussian or Laplacian of Gaussian convolution kernels. We leverage the \emph{Hydra Image Processor} \cite{HIP} for its ability to compute anisotropic scale-space transforms on 3-D images of arbitrary size efficiently using NVIDIA's CUDA architecture. Given a collection of 3-D retinal images, we propose the NCD as a \emph{kernel} for a reproducing kernel Hilbert space \cite{RKHS} that we call the normalized compression vectors (NCV). \textbf{The NCV are an optimally concise and meaningful feature vector representing each input image as a point in a real-valued normalized feature space that reflects the visual input similarities not just at the given image points, but throughout the feature space}. 

We use the NCD with a 3-D generic lossless image compressor called Free Lossless Image Format (FLIF) \cite{FLIF} that uses decision trees to encode spatial patterns in the data. Given $N$ images, we compute the NCD for each pair of images to get an $N \times N$ matrix of pairwise distances. The NCV are defined using classical multidimensional scaling (MDS) \cite{Theodoridis2009} to represent the principal visual differences of the NCD matrix. The NCV features are real-valued, normalized and ranked by order of importance. The NCV preserve the benefits of the reproducing kernel Hilbert space (RKHS) as an equivalency between the metric embedding space and the input images. We combine the NCD with a shape model parameterized by an anisotropic (blob, plate, tube) scale-space \cite{frangi,lindeberg1994scale}. Anisotropic means the voxel physical size is not the same in all three dimensions, a common condition in 3-D optical microscopy. In retinal images, voxels are typically much smaller along the axial dimension compared to the lateral dimensions, optically enhancing the plate-like structures that make up retinal tissue. The NCD is then computed pairwise on the scale-space images and the resulting distances and/or NCV embeddings compared against ground-truth or unsupervised optimization criteria \cite{CSF} to quantify how meaningful the structure is at a given scale. The NCD kernel guarantees equivalence, with smooth gradients and optimality, between the visual (scale-space) and quantitative (NCV space) pattern differences among a collection of images. 

\begin{figure*}[hbt!]
  \includegraphics[width=1.0\textwidth]{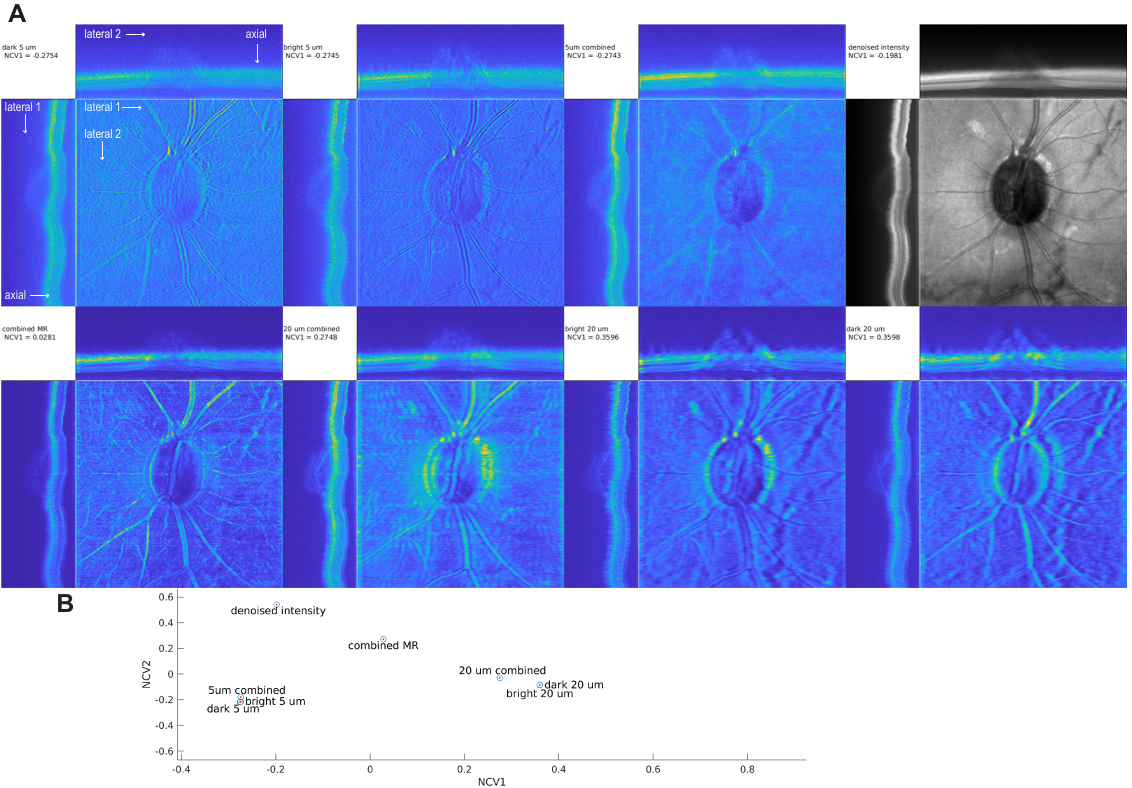}
  \caption[Retinal Structure Function Workflow]{
    \textbf{Anisotropic scale-space combined with a normalized compression distance (NCD) kernel defines an equivalence between visualization (A) and measurement (B) for a single non-human primate 3-D retinal image}. (A) Each of the eight panels shows a "postcard" rendering, divided into three regions showing projections along the three imaging axes. The seven blue postcards show a mean intensity projection of the 3-D structure measures obtained from the parametric (blob, plate, tube)-like scale-space filters. The gray postcard shows the denoised (non-local means) intensity image.  The postcards are sorted by the principal normalized compression vector (NCV1) value of the 3-D image. The 5 $\mu m$ bright and dark filters capture the finest detail while the 20 $\mu m$ filters capture larger structures. The multi-resolution (MR) renderings combine the 5 $\mu m$ and 20 $\mu m$ bright and dark renderings.   (B) We use classical multidimensional scaling on the pairwise NCD matrix to produce the normalized compression vectors (NCV). The NCV features capture the similarity characteristics measured by the NCD kernel throughout the feature space, not just at the points corresponding to input images. \textbf{The NCD kernel defines a universal equivalence between the visual 3-D scale-space images and the quantitative NCV features}.
  }
  \label{fig:nhp_fig1}
\end{figure*}

Figure \ref{fig:nhp_fig1} illustrates the approach. Figure 1A shows a denoised intensity image together with seven structure-enhancing filters (as defined in Section \ref{Sect:scale-space}) applied to a single 3-D retinal image of the optic nerve head of a non-human primate. Note the images are shown in order of increasing NCV1 forming a quantitative visualization of the NCV1 measured pattern differences. Each of the eight images is rendered in a format we term a "postcard". Each postcard rendering consists of 3 2-D projections  showing the whole image stack (not slices) along the different imaging axes. This projection has a "stamp region", the top-left white rectangle formed by the anisotropic corrections for the axial views where we print metadata.  

The first postcard shown in Figure 1A labels the three principal imaging axes (axial, lateral 1, lateral 2). The axial dimension typically covers a smaller physical distance and has more voxels than the lateral dimensions, and the axial projection is often referred to as an "en face" view. For these non-human primate images, the axial dimension (looking into the eye) spans 1.8mm and has 1024 voxels, while each lateral dimension spans 5mm and has 400 voxels. The remaining postcards show the same three projections for different combinations of bright and dark blob and plate filters at different scales. Bright structure is found on the negative response of the scale-space filter, dark structure against a bright background is captured on the positive response. The 5 $\mu m$ filter captures the finest structure for this voxel physical size. The 20 $\mu m$ captures coarser structure \emph{e.g.} axon bundles, vasculature, \emph{etc}.   The panels labeled MR are multi-resolution renderings, combining the 20 and 5 $\mu m$ filters to show fine detail in the context of more significant structure. The combined multi-resolution rendering uses equivalently the sum or max operator between the (non-overlapping) dark and bright structure response and can be observed here as the closest scale-space representation to the (grayscale) input image. The postcards are arranged in order of increasing NCV1, so adjacent panels are most visually similar.  The denoised intensity image is rendered in gray. The scale-space images rendered in blue (parula colormap) are smoother than the raw intensity image, and they preserve more information when projected to 2-D as can be observed by zooming closely to view the 3-D characteristics of the texture. All eight images are compressed separately and together to compute the NCD matrix and the corresponding NCV features. These NCV features are then a quantitative measure of the visual information in the input images. The NCV are sorted by order of importance. The principal NCV1 represents the most informative single feature capturing the pattern differences among the inputs. In theory, any feature that can be computed as a summary statistic of the visual differences in these images can be inferred from the NCV features \cite{Vitanyi2008}. In practice, the NCV are a useful addition to other learning approaches for measuring patterns of change. These 2-D postcard views are the visual representation for the structural differences measured by the NCD kernel.

There are four key contributions to this work. First, we present a novel algorithmic statistics pipeline using lossless image compression for computing automatic visual feature differences on a collection of anisotropic 3-D images. These automatic visual features are called the normalized compression vectors, and in the limit of perfect image compression they define exactly a universal feature space capturing any and all visual pattern differences among the input images.  Second, we validate the pipeline as a measure of structural change among pairs of images against the physician-measured  functional change called visual field mean deviation (VFMD). The NCD shows a strong correlation with the change in VFMD ($\Delta$VFMD) and is a more accurate predictor of $\Delta$VFMD than the current state-of-the-art deep learning approaches. Third, we demonstrate the use of the pipeline to measure the change in structure associated with changes in intraocular pressure settings in non-human primates and to identify the most significant pattern differences among a collection of retinal images from a human with non-progressing glaucoma. Finally, we include a discussion on the impact of non-metric embedding approaches (e.g. neural networks) with a straightforward autoencoder example demonstrating the introduction of class-correlated statistical distortion on output subpopulations \emph{with respect to} the input data.

The remainder of the manuscript is organized as follows. The introduction continues with a description of OCT imaging of the retina, the anisotropic scale-space approach, the distinction between non-metric and algorithmic statistics, and a review of related work. The results section includes the validation of the NCD as a predictor for functional changes of the visual field, and also demonstrates how the NCV features capture patient-specific longitudinal changes in retinal structure. The discussion section includes a summary of the contributions, limitations and future directions. The specific implementations are described in the methods section. The open source software implementations are detailed in the code availability section. 

 \subsection{Optical coherence tomography of the retina}
OCT is a non-invasive diagnostic imaging tool which employs principles of optical interferometry using a low-coherence light source to obtain cross-sectional digitally reconstructed images of biological tissue \cite{huang1991}. OCT images are generated from low coherence interferometry between the tissue reflected light and reference signal \cite{fujimoto2016development}.
OCT has clinical applications in cardiology, otology, dermatology, and dentistry \cite{ali2021optical}.
OCT’s ability to penetrate up to 3mm of tissue and produce high resolution images has provided use cases in diagnosing myocardial infarction with non-obtrusive coronary arteries \cite{reynolds2021coronary}, facilitating cochlear implant surgery \cite{starovoyt2019high}, diagnosing basal cell carcinoma \cite{chen2021evaluation}, and even dental cavity detection \cite{hariri2013estimation}. More recently, the sub-field of ophthalmic imaging known as oculomics has shown the capability of using OCT to make prognostications of non-ocular systemic disorders like cardiovascular \cite{chan2023eyes} and neurological disease \cite{suh2023retina, lin2024individual}, as well as improved estimation of phenotypic age for predicting mortality \cite{nusinovici2022retinal}. The ability to safely and quickly acquire ocular images in a longitudinal and reproducible manner has revolutionized the way clinicians diagnose and manage blindness causing diseases such as glaucoma. Glaucoma is the leading global cause of irreversible blindness, with some projections indicating that the number of affected individuals could rise to 112 million by 2040 due to population aging \cite{tham2014global}. Glaucoma, characterized by persistent structural damage to the optic nerve and retina, is often asymptomatic  until the moderate to severe stages of disease. Therefore, OCT plays a critical role in mitigating disease through early detection of these microscopic structural changes. One of the prominent ways to categorize disease severity and visual field defect is through visual field mean deviation (VFMD), which is a numerical estimation of light in decibels (dB) that an individual eye can perceive \cite{thirunavukarasu2024validated}. All the results in the present manuscript are from 3-D OCT images, but the approach has previously been applied in a number of different live cell and tissue microscopy applications \cite{aho,CSF,Cohen2023}.

\subsection{Visual field measures retinal function}
\label{Sect:Intro:VF}
The Humphrey visual field test \cite{humphrey} uses a dedicated device to measure visual function via light sensitivity of the retina. The visual field mean deviation (VFMD) is a summary measure of the visual field that is commonly used in clinical practice to assess the severity of visual field loss. The VFMD is calculated by comparing the patient's visual field sensitivity to that of a normal reference population, and it is expressed in decibels (dB). With the Humphrey visual field test, a more negative VFMD indicates greater visual field loss. The VFMD is a critical measure for diagnosing and monitoring glaucoma, as well as for assessing the impact of other retinal diseases on visual function. The visual field (VF) test requires patients to sit still and fixate on singular point without moving and click a button anytime a point of light is perceived in their field of vision, which results in high test variability. Therefore, considerable efforts have been made towards predicting VF outcomes based on OCT. In this work, we use changes in VFMD over time as a ground truth measure of functional change to validate the predictive performance of the NCD computed from 3-D OCT images. Although the VFMD is considered the gold standard in assessing retinal function, noisy results often require smoothing for analyzing changes in function over time \cite{abbasi2026comparing}.  The approach proposed for computing visual field here is to calculate the normalized compression distance between pairs of images from the same patient, and then to use a supervised regression model to convert the NCD to a change in VFMD ($\Delta$VFMD). Strictly speaking this value gives the magnitude of the change in $\Delta$VFMD but the typical ophthalmological assumption is that VFMD only decreases over time. If a signed $\Delta$VFMD is required, then the direction of change would need to be established by including additional VFMD labeled images. Improving the ability to predict VFMD changes from OCT images has the potential to improve patient care by providing more accurate assessments of disease progression and treatment efficacy. 

\subsection{Anisotropic scale-space}
\label{Sect:scale-space}
Scale-space is a multi-resolution representation of a parameterized (blob, plate, tube)-like shape model \cite{frangi,lindeberg1994scale}. Scale-space offers a concise visual query for structural content at a given shape surrounding each voxel in the image. The images considered here are 3-D optical coherence tomography retinal images, characterized by large image anisotropy, with much higher axial resolution into the brain, e.g. 2$\mu$m and coarse lateral resolution, e.g. 30$\mu$m. This inherent anisotropy causes significant challenges for structural quantification. Most implementations of the commonly used parameterized shape filters like the Gaussian or Laplacian of Gaussian assume isotropic voxel sizes. As with non-metric embeddings (Section \ref{Sect.non-metric}), using isotropic scale-space filters on anisotropic images can introduce distortion that is difficult to recognize. For the images analyzed here, (Table \ref{tab:table1}) using isotropic scale-space filters on the anisotropic images results in a significantly less accurate relationship between structural changes measured from the images and functional changes measured by VFMD (Table \ref{tab:table2}).

Here we make use of the Laplacian of Gaussian (LoG) imaging filter provided by the Hydra Image Processor (HIP) toolkit \cite{HIP}. The HIP uses NVIDIA CUDA GPUs to implement an efficient metric anisotropic ellipsoidal shape enhancing filter. The filter takes a 3 element covariance (shape) matrix specifying separately the desired radii of the shape space for the axial and both lateral axes. In Section \ref{Sect.VFMD} we show that the NCD computed using the anisotropic scale-space implementation significantly outperforms the standard isotropic scale-space filters for predicting changes in VFMD. The LoG returns bright structure against dark background as negative values, and dark structure on bright background as positive values. Our pipeline processes the bright and dark structure separately, then recombines them using e.g. a max or sum operator in the multi-resolution visualization, as in Figure 1.  

\subsection{Algorithmic  \emph{vs.} non-metric statistics}
For our purposes, an admissible \emph{algorithmic statistic} \cite{Vitanyi2001} is any function whose output is a metric distance computed on the inputs. There are two types of algorithmic statistics used here: metric distances and metric structure functions. Any \emph{metric distance} is admissible \emph{e.g.} the normalized compression distance (NCD) \cite{Vitanyi2005} between pairs of retinal images or the Euclidean distance between pairs of points in an $N-D$ real valued space. Any \emph{metric structure function} is also an admissible algorithmic statistic \cite{CSF,frangi}. A metric structure function, given input data and model parameters, outputs a goodness of fit measure between model and data \cite{Theodoridis2009}. Here we use the class of visual scale-space structure functions, as in \cite{lindeberg1994scale} and \cite{frangi}, with the HIP anisotropic LoG for blob-, plate- and tube- like object enhancement. To be a  metric distance, the axioms are (1) identity $f(X,Y)=0 \iff X=Y$, (2) symmetry $f(X,Y) == f(Y,X)$, and (3) triangle inequality $f(X,Z) \leq f(X,Y) + f(Y,Z)$. To be a metric structure function, the same axioms may be written (1) identity $f(X_i,M)$=0 iff $X_i$ is a perfect match to the model $M$, (2) symmetry $f(X_i,M) == f(X_j,M)$ iff $X_i$ and $X_j$ are equivalently well fit by model $M$, and (3) triangle inequality $|f(X_i,M)-f(X_j,M)| \leq |f(X_i,M)-f(X_k,M)| + |f(X_k,M)-f(X_j,M)|  \forall i,j,k$. The class of functions considered algorithmic statistics is quite broad. Any distance function or any reasonable statistic of the input data is an admissible algorithmic statistic. Similar for structure function, any reasonable measure of goodness of fit between model and data is admissible \cite{KSF,frangi,lindeberg1994scale}. 

A non-metric statistic is a function whose outputs do not obey the axioms of metric distance.
For example, the perceptron function used in neural networks \cite{Theodoridis2009} is a non-metric statistic because it does not obey the triangle inequality.
Non-metric learning machines constitute the vast majority of modern machine learning approaches including deep learning, generative and graphical models.
These machines are powerful and flexible, but their outputs are not interpretable as distances.
The impact of treating non-metric learning machines as metric is highly application dependent, but in Section \ref{Sect.non-metric} we present a simulation using the latent space of a neural network autoencoder to demonstrate how non-metric learning machines can introduce class-correlated statistical distortion on the output subpopulations.

\subsection{Related work} 
The normalized compression distance (NCD) \cite{Vitanyi2005}, is a parameter-free similarity metric grounded in algorithmic information theory that uses generic data compression algorithms to approximate the theoretically perfect normalized information distance \cite{Vitanyi2009}. In the biomedical domain, the NCD has been used to automatically summarize dynamic spatiotemporal changes in biological image sequences \cite{Cohen2009} and also to predict the future differentiation fates of live neural progenitor cells \cite{Cohen2010}.
Recent work has also leveraged NCD and Kolmogorov complexity to construct optimal metric embeddings that capture spatiotemporal cell signaling dynamics in live cell microscopy movies without requiring prior knowledge or training data \cite{aho}. To overcome the theoretical limitations of pairwise comparisons in larger datasets, a multi-set formulation of the NCD \cite{Cohen2015_NCDM} extended the metric to enable  simultaneous multi-object comparisons. The cluster structure function (CSF) \cite{CSF} uses algorithmic statistics to define optimality deficiencies to automatically determine the ideal number of clusters in unsupervised datasets. Spectral clustering \cite{Jordan} utilizes the eigenvectors of an affinity matrix to project complex, non-linear data into a linearly separable lower-dimensional space.
This approach shares theoretical roots with the classical metric multidimensional scaling \cite{Theodoridis2009} used here with the NCD kernel, embedding pairwise distance matrices into Euclidean coordinates via eigenvector decomposition.  Extending these unsupervised foundations, Kamvar et al. \cite{Kamvar} introduced spectral learning for semi-supervised classification, incorporating sparse label data to explicitly adjust Markov transition probabilities within the similarity matrix. Aho et al. \cite{aho} leveraged spectral techniques with the normalized compression distance to construct optimal metric embeddings for classifying live cell microscopy signaling patterns.

The foundation of multi-scale feature extraction was formalized by Lindeberg \cite{lindeberg1994scale} through scale-space theory, which represents image structures across varying resolutions using continuous Gaussian derivative operators. Building on these differential principles, Frangi et al. \cite{frangi} introduced a widely used "vesselness" filter that analyzes the eigenvalues of the multi-scale Hessian matrix to enhance tubular structures. The Frangi vesselness filter implementations available in MATLAB and Python use isotropic Gaussian blob filters that do not properly account for image anisotropy. Using an isotropic filter on an anisotropic image will cause distortion of the image measurement / enhancement. To address this, the Hydra Image Processor (HIP) \cite{HIP}, a GPU-accelerated library that automatically partitions arbitrarily large images across multiple GPUs, was developed. This enables highly parallel spatial filtering operations while utilizing novel kernel renormalization techniques to reduce boundary artifacts. The HIP scales across GPUs and is able to process images of arbitrary size by dividing them into chunks before processing.

The most common approaches to classify glaucoma are based on clinically accepted structural features such as the retinal nerve fiber layer (RNFL), ganglion cell inner plexiform layer (GCIPL) and optic nerve head (ONH) morphology \cite{prahs2018oct, grewal2008artificial, shi2024rnflt2vec, berenguer2021automatic, christopher2018retinal}. Incorporating features such as ganglion cell complex (GCC) thickness, ONH macrostructure, and RNFL reflectance maps into DL model training has been shown to significantly improve diagnostic accuracy \cite{tan2024hybrid}. The approach described here uses the normalized compression vectors (NCV) as a visual feature space that captures any and all differences between the images. Physical features such as RNFL, GCIPL, \emph{etc.} may not be well correlated with the NCV as they measure particular structures within the input images rather than the whole image. Although there is no consensus on the exact clinical classification of glaucoma, the structure function relationship between OCT and VF is critical in understanding disease severity and progression. 

Artificial intelligence (AI) applications using deep learners (DL) to detect, diagnose and predict disease progression from retinal images have been widely used \cite{SCHUMAN2022e3}. Feature agnostic approaches using OCT have shown promise in predicting disease progression measured by VF. Chen et al. \cite{chen2023segmentation}, describe a 3-D based ResNet18 Convolutional neural network (CNN) capable of inferring pointwise VF sensitivities directly from segmentation-free OCT 3-D volumes that achieves a mean error of $\sim$ 3 dB and a correlation coefficient of $\sim$ 0.8. DL approaches using OCT volumes from glaucoma subjects have identified 14 non-clinically defined surface shape patterns at the ONH region capable of predicting specific VFMD loss rate with a squared correlation coefficient of $r^2=0.37$ \cite{saini2022assessing}.  These DL approaches  have demonstrated the ability to identify previously undiscovered biomarkers for improving the prediction of glaucomatous functional defects, but are limited in application and measurement of image characteristics by their underlying non-metric implementations.  A key advantage of the approach presented here is the metric features (NCV and NCD) that can be used to measure the visual differences between images. Non-metric features have been used in some recent works for both retinal imaging and live cell microscopy \cite{aho, MacTel}. Using non-metric features should be avoided as they can introduce systematic distortion that may be correlated with underlying non-meaningful phenotypic differences as described in Section \ref{Sect.non-metric}.

\section{Results}

\subsection{Validating the NCD as a predictor of visual field change}
\label{Sect.VFMD}

To validate the NCD, we compare the change between image pairs, or structural change, against the change in physician-measured visual function ($\Delta$VFMD) using a dataset containing both optic nerve head (ONH) and macula 3-D  volumetric scans. Patient demographics and baseline clinical characteristics are as shown in Table \ref{tab:table1}. The ONH group consisted of 73 subjects (95 eyes), 31 men and 42 women, with average age at earlier scan of 63.3 ± 14.2 years. Over a mean time interval of 20.7 ± 17.9 months, this group yielded 305 right eye (OD) and 339 left eye (OS) OCT image pairs. Clinically, the ONH cohort demonstrated a baseline Visual Field Mean Deviation (VFMD) of -5.21 ± 6.55 dB and a $\Delta$VFMD of -0.81 ± 1.36 dB. The macula group included 51 subjects (60 eyes), consisting of 20 men and 31 women, with a similar mean baseline age of 62.1 ± 13.7 years. This cohort contributed 155 OD and 186 OS OCT pairs evaluated over an average follow-up period of 17.7 ± 12.3 months. The macula group presented with a baseline VFMD of -4.26 ± 5.84 dB and a slightly more negative $\Delta$VFMD of -1.11 ± 2.20 dB compared to the ONH group.

\begin{table*}[h]
    \centering
    \caption{Patient demographics and clinical characteristics. For the number of subjects, total eyes are shown in parentheses: (*) 24 (25.3\%) out of 95 ONH eyes and (**) 14 (23.3\%) out of 60 Macula eyes had a VFMD rate of change less than $-1$ dB/year. Data are presented as mean $\pm$ standard deviation.}
    \begin{tabular}{
        l % 1st column stays naturally sized
        c % ONH data column
        c % Macula data column
    }
        \toprule
        \textbf{Characteristic} & \textbf{ONH} & \textbf{Macula} \\
        \midrule
        {No. of Subjects (Eyes)} & 73 (95*)         & 51 (60**)        \\
        {OCT Pairs (OD/OS)}      & 305/339          & 155/186          \\
        {Sex -- Men/Women}       & 31/42            & 20/31            \\
        {Baseline age (year)}    & $63.3 \pm 14.2$  & $62.1 \pm 13.7$  \\
        {Baseline VFMD (dB)}       & $-5.21 \pm 6.55$ & $-4.26 \pm 5.84$ \\
        {VFMD Change (dB)}       & $-0.81 \pm 1.36$ & $-1.11 \pm 2.20$ \\
        {Time interval (months)}  & $20.7 \pm 17.9$  & $17.7 \pm 12.3$  \\
        \bottomrule
    \end{tabular}
    \label{tab:table1}
\end{table*}

Table \ref{tab:table2} shows the correlation between NCD and $\Delta$VFMD for different scale-space parameters and also for the raw (non scale-space) intensity images. The scale-space filters are anisotropic 3-D metric filters that output bright structure and dark structure separately from the negative and positive Laplacian of Gaussian (LoG) respectively, as detailed in Section \ref{Sect.NCD_methods}. Blob filters were processed at 50 $\mu m$, plate filters used 250 $\mu m$ in both lateral dimensions and 1 $\mu m$ in the axial direction. The sizes were chosen based on voxel physical sizes from the OCT machine, as described in Section \ref{Sect.NCD_methods}. Results are included for an isotropic blob scale-space formulation where the voxel physical size is treated the same in all three dimensions, as with most currently available implementations of the LoG or Gaussian blob filters. The correlation between the NCD and $\Delta$VFMD is not significantly different between the different anisotropic scale-space filter shapes. The NCD derived from bright-blob-filtered macular scan yields the highest correlation coefficient ($r^2 \approx 0.94$). For the ONH scans, the dark plate and raw scan configurations produced the highest correlations ($r^2 \approx 0.92$). NCD from ONH scans yielded slightly lower but highly consistent correlations, ranging tightly between approximately 0.90 and 0.92. 
Table \ref{tab:table2} also shows median absolute error when using the NCD as an input to a Gaussian process regression model for predicting $\Delta$VFMD. Details of the training and prediction for the Gaussian process regression are in Section \ref{Sect.GPR}. In addition to the scale-space and raw intensity image pairwise NCDs, we also consider as a model input the baseline VFMD from the earlier scan date. For both the macula and ONH scans analyzed here, using the baseline VFMD as input to the Gaussian process regression gave a surprisingly accurate prediction with a median absolute error of ~ 0.62 dB for the ONH scans and 0.85 dB for the macula scans. All of the anisotropic scale-space NCD inputs were significantly more accurate ($p<10^{-3}$) compared to the model using only baseline VFMD as input. The NCD computed on inputs processed with the bright blob filter combined with baseline VFMD provided the most accurate prediction, reducing the median absolute error to roughly 0.5 dB for both the ONH and macula datasets. This is lower than the current state-of-the-art model which uses circle and radial scans and yields a mean absolute error of 1.79 dB \cite{Sternetal2023}. These improved error rates generally hovered between 0.55 dB and 0.61 dB for ONH, and 0.68 dB and 0.80 dB for the macula. The isotropic blob filter was significantly less accurate compared to all other models.

\begin{table*}[htbp]
\centering
\caption{\textbf{Performance of normalized compression distance (NCD) and combined with Gaussian Process Regression in predicting changes in Visual Field Mean Deviation ($\Delta$VFMD)}. Table displays the squared correlation coefficient ($r^2$) between NCD and $\Delta$VFMD across distinct image filtering techniques (bright blob, bright plate, dark blob, dark plate, and raw scan). It also shows median absolute error (in dB) of Gaussian process regression models predicting $\Delta$VFMD. The table compares the predictive error when using baseline VFMD alone versus NCD derived from the specified filtered images. Confidence intervals represent 95\% confidence intervals computed from bootstrapping. }
\resizebox{\textwidth}{!}{%
\begin{tabular}{llcccc}
\toprule
 & & \begin{tabular}{@{}c@{}}\textbf{Spearman} \\ \textbf{Correlation}\end{tabular} & \multicolumn{3}{c}{\textbf{Gaussian Regression Model Error}} \\
\cmidrule(lr){3-3} \cmidrule(lr){4-6}
\textbf{Scan Type} & \textbf{Input} & \textbf{r\textsuperscript{2}} & \begin{tabular}{@{}c@{}}\textbf{Median Absolute} \\ \textbf{Error (MedAE)}\end{tabular} & \begin{tabular}{@{}c@{}}\textbf{Confidence Interval} \\ \textbf{(MedAE)}\end{tabular} & \textbf{p-value\textsuperscript{*}} \\
\midrule
ONH & Baseline MD & & 0.619 & [0.586, 0.647] & 0 \\
ONH & Baseline MD \& NCD - bright blob & & 0.493 & [0.434, 0.533] & 6.20E-06 \\
ONH & NCD - bright blob & 0.905 & 0.561 & [0.536, 0.588] & 1.29E-12 \\
ONH & NCD - dark plate & 0.917 & 0.568 & [0.536, 0.594] & 7.38E-29 \\
ONH & NCD - dark blob & 0.904 & 0.569 & [0.540, 0.596] & 1.88E-05 \\
ONH & NCD - bright plate & 0.913 & 0.575 & [0.551, 0.602] & 9.34E-14 \\
ONH & NCD - raw scan & 0.916 & 0.607 & [0.567, 0.635] & 5.10E-30 \\
ONH & NCD - bright blob (isotropic) & 0.909 & 0.724 & [0.694, 0.753] & 3.85E-47 \\
\midrule
macula & Baseline MD & & 0.848 & [0.789, 0.917] & 0 \\
macula & Baseline MD \& NCD - bright blob & & 0.504 & [0.435, 0.542] & 3.23E-15 \\
macula & NCD - dark blob & 0.937 & 0.679 & [0.614, 0.763] & 7.20E-04 \\
macula & NCD - bright plate & 0.929 & 0.753 & [0.710, 0.800] & 1.34E-10 \\
macula & NCD - bright blob & 0.938 & 0.753 & [0.691, 0.825] & 2.34E-05 \\
macula & NCD - dark plate & 0.931 & 0.771 & [0.727, 0.834] & 2.24E-10 \\
macula & NCD - raw scan & 0.926 & 0.798 & [0.740, 0.873] & 6.31E-01 \\
macula & NCD - bright blob (isotropic) & 0.913 & 0.896 & [0.819, 0.940] & 2.67E-05 \\
\bottomrule
\addlinespace
\multicolumn{6}{l}{\textsuperscript{*}\footnotesize p values are computed using the paired Wilcoxon signed-rank test comparing models using NCD as input to the model using just baseline VFMD as input.} \\
\end{tabular}%
}
\label{tab:table2}
\end{table*}

\subsection{Normalized compression vectors measure progression statistics of vision loss}

\begin{figure*}[htb] 
  \centering
  \includegraphics[width=1.0\textwidth]{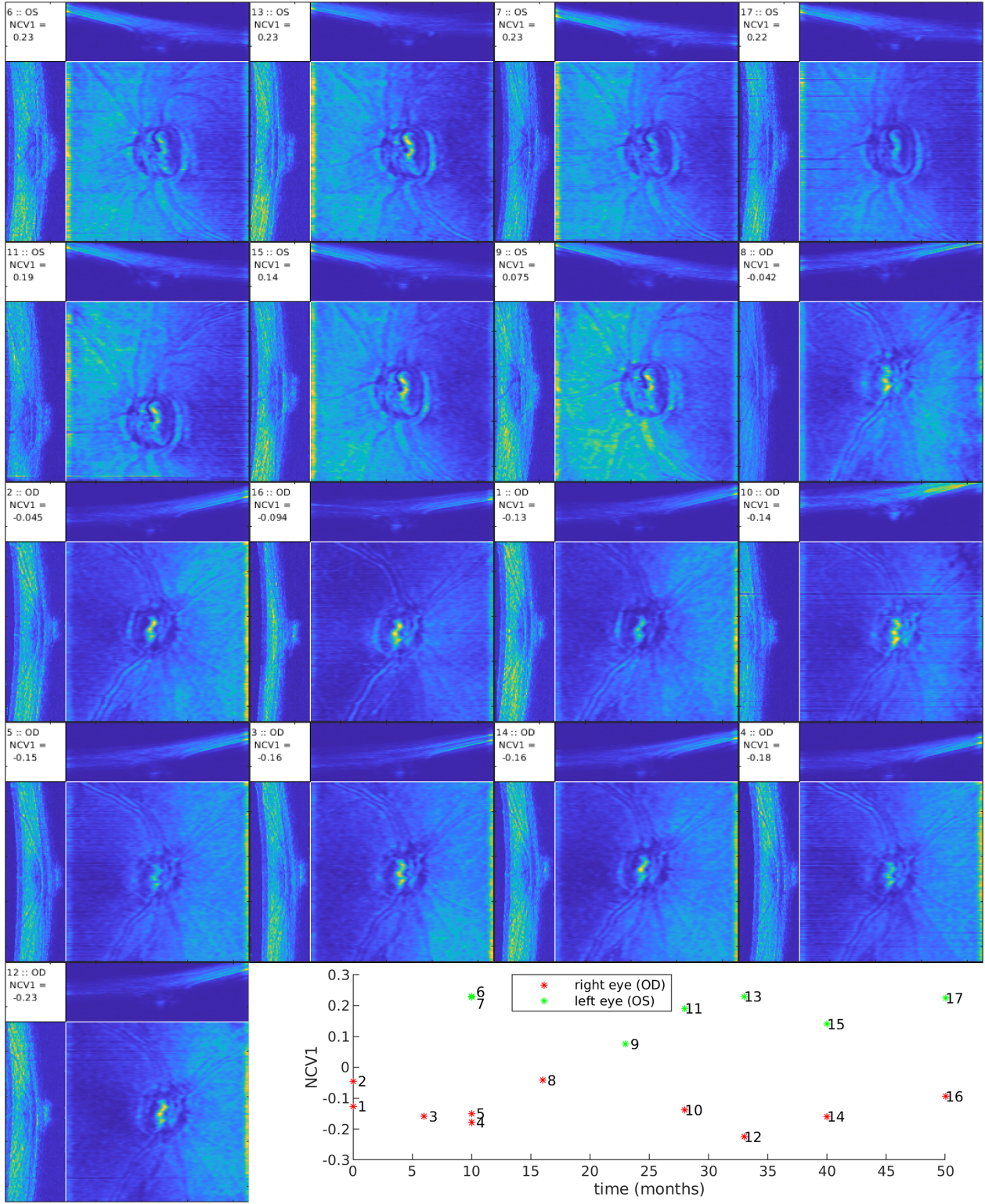}
  \caption[]{
    \textbf{Normalized compression vectors (NCV) capture the pattern differences among a collection of 3-D OCT images of the optic nerve head (ONH) from a patient with non-progressing glaucoma}. The blue scale-space images show multi-resolution renderings of the 3-D structure measurements, sorted by NCV1 values. For this patient, the principal NCV1 captures the two different eyes, left eye (positive) NCV1 vs. right eye (negative) NCV1 as the most interesting single visual difference among the inputs. The trajectories of NCV1 are relatively flat, consistent with the clinical observation that this patient has non-progressing glaucoma. 
  }
  \label{fig:qvHuman}
\end{figure*}

As an example of the ability of the normalized compression distance (NCD) to measure patient-specific longitudinal changes, multidimensional scaling was applied to the NCD dissimilarity matrix for images from a single patient to extract low-dimensional embeddings, termed normalized compression vectors (NCV), 
as detailed in Section \ref{Sect.NCV}. One advantage of the NCV approach is the ability to use the two principal compression vectors as a 2-D visualization of the relationships among the input images.
Figure \ref{fig:qvHuman} shows scale-space images and kernel space embeddings on the bottom right (NCV1 \emph{vs.} time) from a non-progressing subject with both eyes diagnosed with mild glaucoma. For this patient, the biggest difference among the collection of retinal OCT was the left (OS) vs. right (OD) eye. The principal NCV1 captures well-separated clusters for the right and left eyes, with the right eye (OD) having negative NCV1 values and the left eye (OS) having positive values. Visualizing the scale-space images sorted by NCV1 makes the orientation difference that resulted in this NCV feature space visually apparent. In the kernel space embedding scatter plot, neither eye shows a clearly progressive pattern, consistent with the clinical observation that this patient has non-progressing glaucoma. 

\subsection{Relationship between NCD and change in intraocular pressure settings in non-human primates}

We analyzed images from 7 adult macaques, comprising 13 eyes (7 experimental and 6 control; 7 right [OD] and 6 left [OS]). One eye of each macaque underwent laser photocoagulation of the trabecular meshwork to induce chronic intraocular pressure (IOP) elevation and establish an experimental glaucoma model \cite{IOPMethod}. On experimental eyes, images were captured both pre- and post-photocoagulation. The mean age of the cohort was 9.13 ± 1.53 years. The dataset included 132 image pairs (60 OD and 72 OS) captured at 22 imaging sessions. Each image pair consists of two 3-D optic nerve head (ONH) scans (400 $\times$ 400 $\times$ 1024 voxels, 5 $\times$ 5 $\times$ 1.8 mm) of the same eye and session taken at 3 different pressure settings (5, 15, and 30 mmHg) to evaluate the effects of changing pressure setting. Of the 132 imaging pairs, 48 pairs were from control eyes, 42 pairs were from experimental eyes that had not undergone laser photocoagulation procedure, and 42 pairs were post procedure. Using the Laplacian of Gaussian (LoG) anisotropic scale-space filter the NCD was computed for each image pair and correlated with the change in pressure setting $\Delta$IOP between the two scans. For this application, the scale-space filters were run for 5 $\mu m$ blobs, the minimum blob size for the voxel physical size of these images. The pressure setting change was calculated as the difference in pressure setting between the two scans, with possible values of 0 mmHg (same pressure), 10 mmHg (5 to 15), 15 mmHg (15 to 30), or 25 mmHg (5 to 30). Across all 132 image pairs, the squared correlation coefficient between NCD and pressure setting change was $r^2 = 0.67 $ for bright and $r^2 = 0.64 $ for dark LoG response. When comparing the 42 image pairs pre-photocoagulation with the 42 pairs post-photocoagulation for the experimental group as shown in Table \ref{tab:table3}, the correlation between NCD and pressure setting change was not statistically significantly different.  Different eyes exhibited different amounts of structural change in response to pressure settings. This may be due to inherent physiological characteristics, or it could be an effect of the relatively small sample size compared to the human data analyzed in Section \ref{Sect.VFMD}.

 \begin{table*}[h]
     \centering
     \caption{Squared Spearman correlation between NCD and change in intraocular pressure ($\Delta$IOP) setting for non-human primates. For each animal on a given day, retinal images were acquired at three different intraocular pressure settings. Laser treatment induces an experimental glaucoma model. The NCD was computed between each image pair at different pressure settings from the same eye on the same date. }
     \begin{tabular}{
         l % 1st column stays naturally sized
         S[table-format=1.2, table-column-width=1.5cm] % r^2
         S[table-format=1.1e2, table-column-width=2cm] % p-value
         S[table-format=1.2, table-column-width=1.5cm] % r^2
         S[table-format=1.1e2, table-column-width=2cm] % p-value
         S[table-format=1.2, table-column-width=1.5cm] % r^2
         S[table-format=1.1e2, table-column-width=2cm] % p-value
     }
         \toprule
         & \multicolumn{2}{c}{\textbf{Entire Cohort}} & \multicolumn{2}{c}{\textbf{Before Laser}} & \multicolumn{2}{c}{\textbf{After Laser}} \\
         \cmidrule(lr){2-3} \cmidrule(lr){4-5} \cmidrule(lr){6-7}
         \textbf{Radii} & {$\mathbf{r^2}$} & {\textbf{p-value}} & {$\mathbf{r^2}$} & {\textbf{p-value}} & {$\mathbf{r^2}$} & {\textbf{p-value}} \\
         \midrule
         {bright blob}    & 0.67 & 4.5e-33 & 0.66 & 6.7e-11 & 0.66 & 1.2e-10 \\
         {dark blob}      & 0.64 & 4.6e-31 & 0.59 & 2.6e-09 & 0.61 & 8.2e-10 \\
         \bottomrule
     \end{tabular}
     \label{tab:table3}
 \end{table*}

\subsection{Statistical distortion from non-metric features}
\label{Sect.non-metric}
Metric embeddings preserve distance relationships among the input data. For example, given an input consisting of points from two Gaussian distributions with different mean values and identical variance, a metric embedding  will result in both populations having the same variance in the embedding or feature space. Non-metric embeddings do not preserve distance relationships, so variance in the embedded feature space may be different for the two populations. The impact of a non-metric embedding will vary by application. Here we consider a straightforward example. To evaluate distortion in a neural network latent space embedding, synthetic datasets comprising two 2-D Gaussian distributed populations with different mean and identical standard deviation ($n = 1000$, $\sigma = 1$, $\mu = \pm 1$)  were generated. The concatenated data was embedded using two methods: a shallow autoencoder built from a single-layer neural network and classical metric multidimensional scaling (MDS) \cite{Theodoridis2009}. Synthetic data was generated iteratively, and the variance of the data in the embedding space was computed. Paired-sample $t$-tests were conducted for 100 samples of the embedding variances for each population on the input data, and also using both MDS and the autoencoder. The input data variance and the MDS embedding variance showed no significant differences between the two populations ($p=0.95$). The autoencoder latent space embedding showed a significant difference in variance between the two populations ($p<10^{-100}$). Mathematically, the interpretation of the autoencoder latent space is that the Euclidean distance between embedding points does not exist. In general, non-metric feature spaces or embeddings should not be used for inference or measurement.

\section{Discussion \& Conclusions}
\label{Sect:discussion}

The normalized compression distance (NCD) kernel combined with anisotropic scale-space visualizations provides a toolbox for optimally visualizing and measuring pattern differences among a collection of images. Using the NCD as a kernel for a reproducing Hilbert feature space called the normalized compression vectors enables the measurement of different visual representations of the underlying image structures. The present work validates the NCD as a robust, parameter-free metric for quantifying 3-D OCT retinal changes and accurately predicting functional vision loss measured by visual field mean deviation (VFMD). Combining the NCD with anisotropic scale-space filters gives an efficient pipeline for visualizing and measuring structural changes among a collection of images. Because OCT images possess inherent anisotropy, characterized by different axial and lateral resolutions, standard isotropic filters do not accurately measure spatial structure.  Multidimensional scaling of the NCD dissimilarity matrix yields normalized compression vectors (NCV), a concise, low-dimensional embedding for measuring longitudinal progression.  In a non-human primate model this approach was used to measure structural change induced by intraocular pressure setting variations, finding that changes between image pairs are less correlated with changes in intraocular pressure than with changes in visual field function of the eye as measured in the human population. The NCD is able to effectively quantify visually apparent structural differences in 3-D between image pairs. Small training set supervised regression models were trained on the NCD output space to map the NCD to $\Delta$VFMD, leveraging the smooth characteristics of the metric embedding (RKHS) kernel to improve the generalizability and accuracy of the approach. 

Non-metric embeddings introduce class-correlated statistical distortions that fail to preserve true geometric distance relationships. Measuring microscopy images for biomedical applications requires that differences are measured accurately and reflect genuine anatomical progression rather than inherent algorithmic distortion. Our approach to predicting the change in visual field function ($\Delta$VFMD) differs from most existing approaches but is able to achieve improved accuracy with a fundamentally unsupervised model. One challenge with the metric learning and features proposed here is the challenge of interpreting the specific structural changes that are being captured by the NCD. The NCD is a global measure of the difference between two images, and does not provide information about which specific features or regions of the image are driving the differences. Given the whole image as input, the NCD will not generally capture changes in a specific structure, \emph{e.g.} the retinal nerve fiber layer (RNFL), but rather will measure differences across the entirety of input images. Future work will focus on developing methods to localize and interpret the specific structural changes that are being captured by the NCD, which could provide valuable insights into the underlying disease processes and improve the clinical utility of this approach.

\section{Methods}
\subsection {Retinal Imaging}

OCT images from human ONH and macula were used for the validation of NCD against visual field (see Section \ref{Sect.VFMD}).
3-D images were obtained using Spectral Domain OCT (SD-OCT; Cirrus-HD OCT; Zeiss, Dublin, CA). Glaucoma subjects were enrolled from our ongoing prospective, longitudinal study designed to assess ocular structure over time. The institutional review boards and ethics committees at New York University approved the study (IRB number 16-01302). The study followed the tenets of the Declaration of Helsinki and was conducted in compliance with the Health Insurance Portability and Accountability Act (HIPAA). Informed consent was obtained from all subjects. 
Each image pair comprised a baseline and a corresponding follow-up scan of the same eye.
Both OCT imaging and visual field assessments were conducted concurrently on the same day during the baseline and follow-up visits. To be included in the analysis, images were required to be volumetric OCT scans with a resolution of 200 $\times$ 200 $\times$ 1024 voxels, demonstrate adequate signal strength $\geq$ 6 without significant motion artifacts, and be acquired from eyes with no concurrent ophthalmic pathologies that could alter retinal appearance. Visual field mean deviation (VFMD) was evaluated using the Humphrey Field Analyzer equipped with the Swedish Interactive Thresholding Algorithm (SITA) Standard 24-2 program (Zeiss, Dublin, CA). Due to high variability of VFMD measurements, a linear fit model was applied to smooth the measurement over time as in \cite{abbasi2026comparing}.

In the non-human primate dataset, we analyzed 3-D optic nerve head (ONH) scans from 7 adult macaques (3 rhesus (1 male, 2 females) and 4 cynomolgus (4 males) macaques). All animal procedures were reviewed and approved by the Institutional Animal Care and Use Committee (IACUC) of State University of New York Downstate Health Sciences University. The study adhered to the National Institutes of Health Guide for the Care and Use of Laboratory Animals and to the Association for Research in Vision and Ophthalmology statement for the use of animals in ophthalmic and vision research. Experiment preparation was as described previously \cite{IOPMethod}. Briefly, the primates were placed prone with their heads upright and anesthetized using ketamine and xylazine. To test varying pressure conditions, intraocular pressure (IOP) was controlled using a gravity-based saline perfusion system connected to the anterior chamber through a corneal cannulation. During the experiment, the IOP was modulated through various pressure settings (5, 15, or 30mmHg), allowing a 5-minute stabilization period before capturing OCT scans of the optic nerve head at each stage. For every IOP setting, the optic nerve head (ONH) was captured using a spectral-domain OCT device (Envisu; Leica, Chicago, IL) operating at 20,000 A-scans per second, yielding a 5 $\times$ 5 $\times$ 1.8 mm volumetric scan.

\subsection{Computing normalized compression distance}
\label{Sect.NCD_methods}
The initial image processing pipeline starts with denoising via a three-dimensional (3-D) non-local means algorithm \cite{HIP}. Following denoising, the 3-D volumes are processed using anisotropic 3-D Laplacian of Gaussian (LoG) scale-space filters \cite{HIP} to enhance features within optic nerve head (ONH) and macular structures. The human and non-human primate OCT scans are highly anisotropic with a physical voxel size of $[30, 30, 2]$ $\mu m/voxel$ and $[12.5, 12.5, 1.76]$ $\mu m/voxel$, respectively. The first two dimensions refer to the lateral (low resolution) and the third dimension refers to the axial (high resolution looking into the eye). The LoG filter captures bright structure against a dark background as a negative response, and dark structure against a bright background as a positive response. 

To effectively capture structures across different scan types and experimental needs, specific LoG filter radii were applied based on the computational limits of the Hydra Image Processor (HIP) and empirical observations. For the non-human primate scans depicted in Figure \ref{fig:nhp_fig1}, the LoG filters were set to 5 $\mu m$ and 20 $\mu m$, representing 0.4 $\times$ max(physical voxel size) and 4 $\times$ 0.4 $\times$ max(physical voxel size), respectively. The baseline radius of 0.4 $\times$ max(physical voxel size) was selected because it represents the highest resolution of the LoG filter that the HIP can compute. Conversely, the human images shown in Figure \ref{fig:qvHuman} utilized a multi-resolution LoG with blob radii of 12 $\mu m$ (0.4 $\times$ max(physical voxel size)) and 24 $\mu m$ (2 $\times$ 0.4 $\times$ max(physical voxel size)). However, to predict the change in visual field mean deviation (VFMD) using the normalized compression distance (NCD) in Table \ref{tab:table2}, human images were computed at different scales to match the morphology of the eye. Both blob-like and plate-like kernel sizes were compared. The blob filters use a 50 $\mu m$ radius, and the plate filters use a size of [250, 250, 1] $\mu m$. These specific dimensions are optimally tuned to retinal anatomy and the physical resolution limits of OCT imaging. The plate filter's 1 $\mu m$ axial dimension aligns with the extremely fine axial resolution of the scans, preventing the blurring of thin horizontal boundaries between distinct retinal layers. Meanwhile, its broad 250 $\mu m$ lateral span captures the macroscopic continuity of these planar tissue sheets, which are optically enhanced as plate-like structures by the inherent voxel anisotropy. Conversely, the 50 $\mu m$ blob radius targets structures with roughly a 100 $\mu m$ diameter, such as major retinal blood vessels and thick axon bundles. This specific scale allows the filter to highlight critical functional biomarkers while effectively ignoring high-frequency speckle noise and remaining distinct from the broad, flat layers managed by the plate filter.

Additionally, to directly compare the efficacy of accounting for true scan dimensions, an isotropic testing case was generated by setting the physical voxel size to an isotropic $[30, 30, 30]$ $\mu m/voxel$ during the bright blob filtering process. After conversion to scale-space, the images are quantized to 8 bits. The quantization considers the positive (dark) and negative (bright) response of the LoG separately. The images are quantized into an 8-bit unsigned integer space by linearly mapping the LoG intensities to an 8-bit voxel value $[0,255]$. After quantization, image compression uses the Free Lossless Image Format (FLIF) \cite{FLIF}. FLIF uses a decision tree encoding framework to identify spatial patterns in the data and achieves excellent compression for 3-D image stacks \cite{CSF}. The 3-D OCT volumes are compressed individually to determine their respective file sizes in bytes, denoted as $C(x)$ and $C(y)$. The paired images are then concatenated and compressed together to yield the joint compressed size, $C(xy)$. Images are concatenated vertically and horizontally w.r.t. the high resolution axis and the smaller NCD is chosen. The structural difference between the two images is computed using the normalized compression distance (NCD), as defined in Equation \ref{eq:ncd_formula}.

\begin{equation}
\label{eq:ncd_formula}
NCD(x, y) = \frac{C(xy) - \min\{C(x), C(y)\}}{\max\{C(x), C(y)\}}
\end{equation}

\subsection{Prediction Modeling in the RKHS embedding space}
\label{Sect.GPR}
To convert the NCD value between two Optical Coherence Tomography (OCT) images to an equivalent change in Visual Field Mean Deviation ($\Delta$VFMD), we use a Gaussian process regression (GPR) \cite{GPR}. Training the supervised regression model on the NCD kernel output values simplifies the learning machine compared to training directly on the input images. The GPR works by defining a prior distribution over functions and using observed training data to update this into a predictive posterior distribution. Model performance was evaluated using the Median Absolute Error (MedAE) in decibels (dB). The GPR model was trained and evaluated using leave-one-out-cross validation on several distinct input feature sets to compare the predictive utility of clinically measured visual function against structural changes measured by the NCD. For each pair of images, baseline VFMD is defined by the VFMD measured at the earlier image date. Input configurations included using baseline VFMD alone, NCD derived from raw OCT scans, and NCD extracted via Laplacian of Gaussian (LoG) filters designed to isolate specific geometries such as bright or dark blobs and plates. To quantify the impact of using anisotropic voxel dimensions, an additional model utilized NCD from bright-blob isotropic filtered images with the physical voxel size standardized to $[30, 30, 30]$ $\mu m$/voxel. A GPR model was also trained using both baseline VFMD and NCD derived from bright-blob filtered images to assess the combined predictive power of functional and structural data. All training data was curated with a strict, non-overlapping split protocol to ensure absolute independence between the training and testing sets.

\subsection{Metric Embedding}
\label{Sect.NCV}
To visualize and evaluate longitudinal structural changes in patient eyes the pairwise structural dissimilarities between all scans for a given patient were quantified by computing the pairwise normalized compression distance (NCD) matrix. This matrix is symmetric with zeros on the diagonal. To translate these pairwise relationships into an interpretable geometric space, classical multidimensional scaling (MDS) was applied to the NCD matrix. MDS converts an $N \times N$ distance matrix into an $N$-dimensional feature space, the normalized compression vectors (NCV) that exactly preserve the pairwise distance relationships between the input images. Because the NCD is a metric distance, the NCV features define a reproducing kernel Hilbert space, a type of feature space known for accurate measurement, visualization and inference \cite{RKHS}. The NCV features are ranked by order of importance, with NCV1 capturing the largest direction of pattern difference among the input images. To visualize the NCV, we extract the first two dimensions of this projection (NCV1 and NCV2). These principal NCV were subsequently plotted against time (months from first image in the dataset) and against one another. This technique allows for the visual assessment of distinct temporal trajectories, highlighting visually apparent morphological pattern differences.

\subsection{Optimality of the normalized compression vectors}

The normalized compression vectors are automatic \emph{visual features} defined by the normalized compression distance (NCD) kernel. The NCD is optimal in that it \emph{"minorizes every computable similarity distance up to an error that depends on the quality of the compressor’s approximation of the true Kolmogorov complexities"} \cite{Vitanyi2005}. Better compression means more accurate distance and \emph{vice versa}. The NCD uses concepts from Kolmogorov complexity theory to measure the most concise and meaningful description of any and all differences between pairs of images \cite{CSF,Cohen2015_NCDM}.  Given a collection of $N$ input images $X = \{x_1,x_2,...,x_N\}$, the pairwise NCD matrix is an $N \times N$ square matrix where the entry at position $(i,j)$ represents the NCD between images $x_i$ and $x_j$. The NCD matrix $NCD(X)$ is symmetric and has zeros on the diagonal, guaranteeing real eigenvalues and enabling the use of classical metric multidimensional scaling or spectral clustering \cite{Theodoridis2009} to extract the principal patterns of difference among the input images. 

The normalized compression vectors presented in this work are computed by applying classical metric multidimensional scaling (MDS) to the pairwise NCD matrix $NCD(X)$. We write $Y = MDS(NCD(X))$, where $Y$ is the resulting $N \times N$ matrix of normalized compression vectors (NCV). The input images $X$ are \emph{embedded} in feature space $Y$. The $i^{th}$ row of $Y$ corresponds to the NCV for image $x_i$, the columns of $Y$ are the NCV features.  The NCV are ranked by order of importance, with NCV1 capturing the largest pattern difference among the input images. If all $N$ NCV are preserved, then $Y$ exactly preserves the pairwise distance relationships between all input images. If only the first $K$ NCV are preserved the resulting $Y_K$ optimally preserves the pairwise distance relationships. The NCD is a reproducing kernel for a Hilbert space \cite{RKHS}. This means that the NCV features preserve the visual similarity characteristics not just at the points corresponding to the input images, but also at all points throughout the feature space. This is a key advantage of the NCD approach, as it allows for accurate measurement and inference even for images that were not part of the original input set. 

The NCD will always return the same distance for a given pair of images, regardless of the context of the other images in the collection of input images. The NCV however is defined for a particular collection of images. Given two collections of images $X_1$ and $X_2$ that share some images in common, the NCD between any two images that are in both collections will be the same regardless of whether it is computed as part of $X_1$ or $X_2$. The NCV however will be different for the same image when computed as $NCV(X_1)$ versus $NCV(X_2)$, because the NCV is defined by the pairwise relationships among all images in the collection. This means that the NCV is not a fixed property of an individual image, but rather a property of the image in relation to the other images in the collection. 

\section{Hardware, software and data availability}

All of the software tools used are available free and open source, see \url{https://git-bioimage.coe.drexel.edu/opensource/ncd}. This repository includes the source code and image used to generate Figure 1. For the statistical distortion introduced by non-metric embeddings, source code is available at \url{https://git-bioimage.coe.drexel.edu/opensource/inherent-bias}. The reference hardware platform uses 128 CPU cores (AMD 7763), 2 NVIDIA A6000 GPU, 2TB RAM, 1PB HDD. 

\label{Sect.Software}

\section{Acknowledgements}
Portions of this work were supported by grants from the National Eye Institute of the National Institutes of Health (NIH R01-EY013178, NIH R01-EY030770, and NIH R01-EY035174).

\printbibliography

\newpage
 
\end{document}